\documentclass[review]{elsarticle}

\usepackage{lineno,hyperref}
\usepackage[colorinlistoftodos,prependcaption,textsize=tiny]{todonotes}
\usepackage{multirow}
\modulolinenumbers[5]

\usepackage{color}

\journal{Computer \& Security}

\usepackage[utf8]{inputenc}

\usepackage{adjustbox}
\usepackage{graphicx}
\usepackage{amssymb}
\usepackage{rotating}
\usepackage{tabularx}

\newcommand\tabrotate[1]{\begin{turn}{30}\rlap{#1}\end{turn}}

\begin{document}

\begin{frontmatter}

\title{Flow-based Network Traffic Generation using Generative Adversarial Networks}

\author[co]{Markus Ring\corref{mycorrespondingauthor}}
\cortext[mycorrespondingauthor]{Corresponding author}
\ead{markus.ring@hs-coburg.de}

\author[wue]{Daniel Schlör}
\ead{daniel.schloer@informatik.uni-wuerzburg.de}

\author[co]{Dieter Landes}
\ead{dieter.landes@hs-coburg.de}

\author[wue]{Andreas Hotho}
\ead{hotho@informatik.uni-wuerzburg.de}

\address[co]{Faculty of Electrical Engineering and Informatics, Coburg University of Applied Sciences, 96450 Coburg, Germany}
\address[wue]{Data Mining and Information Retrieval Group, University of Würzburg, 97074 Würzburg, Germany}

\begin{abstract}
Flow-based data sets are necessary for evaluating network-based intrusion detection systems (NIDS).
In this work, we propose a novel methodology for generating realistic flow-based network traffic.  
Our approach is based on Generative Adversarial Networks (GANs) which achieve good results for image generation. 
A major challenge lies in the fact that GANs can only process continuous attributes. However, flow-based data inevitably contain categorical attributes such as IP addresses or port numbers.
Therefore, we propose three different preprocessing approaches for flow-based data in order to transform them into continuous values.
Further, we present a new method for evaluating the generated flow-based network traffic which uses domain knowledge to define quality tests. 
We use the three approaches for generating flow-based network traffic based on the CIDDS-001 data set. 
Experiments indicate that two of the three approaches are able to generate high quality data. 

\end{abstract}

\end{frontmatter}

\section{Introduction}
\label{sec:introduction}
Detecting attacks within network-based traffic has been of great interest in the data mining community over decades.
Recently, Buczak and Guven~\cite{buczak2016survey} presented an overview of the community effort with regard to this issue. 
However, there are still open challenges (e.g. the high cost of false-positives or the lack of  labeled data sets which are publicly available) for the successful use of data mining algorithms for anomaly-based intrusion detection \cite{sommer2010outside,catania2012automatic}. 
In this work, we focus on a specific challenge within that setting.

\textbf{Problem Statement.}
For network-based intrusion detection (NIDS), few labeled data sets are publicly available which contain realistic user behavior and up-to-date attack scenarios. 
Available data sets are often outdated or suffer from other shortcomings. 
Using real network traffic is also problematic due to the missing ground truth. 
Since flow-based data sets contain millions up to billions of flows, manual labeling of real network traffic is difficult even for security experts and extremely time-consuming. 
As another disadvantage, real network traffic often cannot be shared within the community due to privacy concerns. 
However, labeled data sets are necessary for training supervised data mining methods (e.g. classification algorithms) and provide the basis for evaluating the performance of supervised as well as unsupervised data mining algorithms.

\textbf{Objective.}
Large training data sets with high variance can increase the robustness of anomaly-based intrusion detection methods. 
Therefore, we intend to build a generative model which allows us to generate realistic flow-based network traffic. 
The generated data can be used to improve the training of anomaly-based intrusion detection methods as well as for their evaluation. 
To that end, we propose an approach that is able to learn the characteristics of collected network traffic and generates new flow-based network traffic with the same underlying characteristics.

\textbf{Approach and Contributions.} 
Generative Adversarial Networks (GANs)~\cite{goodfellow2014generative} are a popular method to generate synthetic data by learning from a given set of input data. 
GANs consist of two networks, a generator network $G$ and a discriminator network $D$. 
The generator network $G$ is trained to generate synthetic data from noise. 
The discriminator network $D$ is trained to distinguish generated synthetic data from real world data. 
The generator network $G$ is trained by the output signal gradient of the discriminator network $D$. 
$G$ and $D$ are trained iteratively until the generator network $G$ is able to fool the discriminator network $D$. 
GANs achieve remarkably good results in image generation \cite{radford2015unsupervised,ledig2016photo,isola2017image,arjovsky2017wasserstein}.
Furthermore, GANs have also been used for generating text \cite{yu2017seqgan} or molecules~\cite{preuer2018fr}. 

This work uses GANs to generate complete flow-based network traffic with all typical attributes. 
To the best of our knowledge, this is the first work that uses GANs for this purpose. 
GANs can only process continuous input attributes. 
This poses a major challenge since flow-based network data consist of continuous and categorical attributes. 
Consequently, we analyze different preprocessing strategies to transform categorical attributes of flow-based network data into continuous attributes. 
The first method simply treats attributes like \textit{IP addresses} and \textit{ports} as numerical values. 
The second method creates binary attributes from categorical attributes.  
The third method uses IP2Vec~\cite{ring2017ip2vec} to learn meaningful vector representations of categorical attributes. 
After preprocessing, we use Improved Wasserstein GANs (WGAN-GP)~\cite{gulrajani2017improved} with the two time-scale update rule (TTUR) proposed by Heusel et al.~\cite{heusel2017gans} to generate new flow-based network data based on the public CIDDS-001~\cite{ring2017flow} data set.
Then, we evaluate the quality of the generated data with several evaluation measures. 

The paper has several contributions. 
The main contribution is the generation of flow-based network data using GANs. 
We propose three different preprocessing approaches and a new evaluation method which uses domain knowledge to evaluate the quality of generated data. 
In addition to that, we extend IP2Vec \cite{ring2017ip2vec} such that IP2Vec is able to learn similarities between the flow attributes: \textit{bytes}, \textit{packets} and \textit{duration}.

\textbf{Structure.} 
The next section of the paper describes flow-based network traffic, GANs, and IP2Vec in more detail. 
In section~\ref{sec:approach}, we present three different approaches for transforming flow-based network data. 
An experimental evaluation of these approaches is given in section~\ref{sec:experiments} and the results are discussed in section~\ref{sec:discussion}. 
Section~\ref{sec:relatedwork} analyzes related work on network traffic generators for flow-based data. 
A summary and outlook on future work concludes the paper.

\section{Foundations}
\label{sec:basics}
This section starts with analyzing the underlying flow-based network traffic. 
Then, GANs are explained in more detail. 
Finally, we explain IP2Vec \cite{ring2017ip2vec} which is the basis of our third data transformation approach.  

\subsection{Flow-based Network Traffic}
We focus on flow-based network traffic in unidirectional \textit{NetFlow} format~\cite{claise2004cisco}. 
Flows contain header information about network connections between two endpoint devices like servers, workstation computers or mobile phones. 
Each flow is an aggregation of transmitted network packets which share some properties~\cite{claise2008specification}.
Typically, all transmitted network packets with the same \textit{source IP address}, \textit{source port}, \textit{destination IP address}, \textit{destination port} and \textit{transport protocol} within a time window
are aggregated into one flow.
\textit{NetFlow} \cite{claise2004cisco} aggregates all network packets which share these five properties into one flow until an active or inactive timeout is reached.  
In order to consolidate contiguous streams the aggregation of network packets stops if no further packet is received within a time window of $\alpha$ second  (inactive timeout). 
The active timeout stops the aggregation of network packets after $\beta$ seconds, even if further network packets are observed to avoid unlikely long entries. 

\begin{table}[]
	
	\caption{
		Overview of typical attributes in flow-based data like \textit{NetFlow}~\cite{claise2004cisco} or \textit{IPFIX}~\cite{claise2008specification}.
		The third column provides the type of the attributes and the last column shows exemplary values for these attributes. 
	}
	
	\label{tbl:netflow}
	\centering
	\begin{tabular}{p{1em}p{10.5em}p{8em}p{10.5em}}
		\hline\noalign{\smallskip}
		\# & Attribute & Type & Example \\
		\hline
		1 & date first seen & timestamp & 2018-03-13 12:32:30.383 \\
		2 & duration & continuous & 0.212 \\
		3 & transport protocol & categorical & TCP \\
		4 & source IP address & categorical & 192.168.100.5 \\
		5 & source port & categorical & 52128 \\
		6 & destination IP address & categorical & 8.8.8.8 \\
		7 & destination port & categorical & 80 \\
		8 & bytes & numeric & 2391\\
		9 & packets & numeric & 12 \\
		10 & TCP flags & binar/categorical & .A..S. \\
	\end{tabular}
	
\end{table}

Table \ref{tbl:netflow} shows the typical attributes of unidirectional \textit{NetFlow} \cite{claise2004cisco} data. 
As shown in Table \ref{tbl:netflow}, \textit{NetFlow} are heterogeneous data which consists of continuous, numeric, categorical and binary attributes. 
Most attributes like \textit{IP addresses} and \textit{ports} are categorical. 
Further, there is a timestamp attribute (\textit{date first seen}), a continuous attribute (\textit{duration}) and numeric attributes like \textit{bytes} or \textit{packets}. 
We defined the type of \textit{TCP flags} as binary/categorical. 
\textit{TCP flags} can be either interpreted as six binary attributes (e.g. isSYN flag, isACK flag, etc.) or as one categorical value. 

\subsection{GANs}
\begin{figure}[]
	\centering
	\includegraphics[width=0.6\textwidth]{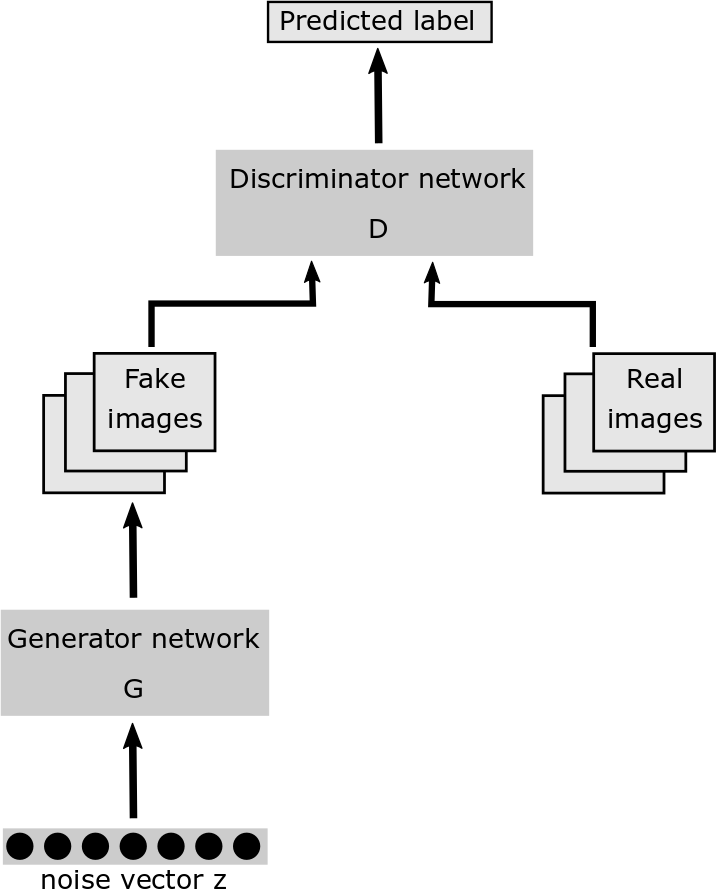}
	\caption{ Architecture of GANs. }
	\label{fig:overviewgan}
\end{figure}

Discriminative models classify objects into predefined classes \cite{han2011data} and are often used for intrusion detection (e.g. in \cite{beigi2014towards}, \cite{stevanovic2015analysis} or \cite{wagner2011machine}). 
In contrast to discriminative models, generative models are used to generate data like flow-based network traffic.
Many generative models build on likelihood maximization for a parametric probability distribution. 
As the likelihood function is often unknown or the likelihood gradient is computationally intractable, some models like Deep Boltzmann Machines \cite{salakhutdinov2010efficient} use approximations to solve this problem. 
Other models avoid this problem by not explicitly representing likelihood. 
Generative Stochastic Networks for example learn the transition operation of a Markov chain whose stationary distribution estimates the data distribution.
GANs avoid Markov chains estimating the data distribution by a game-theoretic approach: 
The generator network $G$ tries to mimic samples from the data distribution, while the discriminator network $D$ has to differentiate real and generated samples.
Both networks are trained iteratively until the discriminator $D$ can't distinguish real samples from generated samples any more. 
Beside computational advantages, the generator $G$ is never updated with real samples. 
Instead, the generator network $G$ is fed with an input vector of noise $z$. 
The generator is trained using only the discriminator's gradients through backpropagation. 
Therefore, it is less likely to overfit the generator $G$ by memorization and reproduction of real samples. 
Figure \ref{fig:overviewgan} illustrates the generation process. 

Goodfellow et al. say that: \textit{"another advantage of adversarial networks is that they can represent very sharp, even degenerate distributions"} \cite{goodfellow2014generative} which is the case for some \textit{NetFlow} attributes. 
However, (the original) vanilla GANs \cite{goodfellow2014generative} require the visible units to be differentiable, which is not the case for categorical attributes like \textit{IP addresses} in \textit{NetFlow} data. 
Gulrajani et al. \cite{gulrajani2017improved} show that Wasserstein GANs (WGANs), besides other advantages, are capable of modeling discrete distributions over a continuous latent space.
In contrast to vanilla GANs, WGANs \citep{arjovsky2017wasserstein} use the Earth Mover (EM) distance as a value function replacing the classifying discriminator network with a \textit{critic} network estimating the EM distance.
While the original WGAN approach uses weight clipping to guarantee differentiability almost everywhere, Gulrajani et al.  \cite{gulrajani2017improved} improve training of WGANs by using gradient penalty as soft constrain to enforce the Lipschitz constraint.
One research frontier in the area of GANs is to solve the issue of non-convergence~\cite{goodfellow2016nips}.
Heusel et al.~\cite{heusel2017gans} propose a two time-scale update rule (TTUR) for training GANs with arbitrary loss functions.
The authors prove that TTUR converges under mild assumptions to a stationary local Nash equilibrium. 

For those reasons, we use Improved Wasserstein Generative Adversarial Networks (WGAN-GP)~\cite{gulrajani2017improved} with the two time-scale update rule (TTUR) from~\cite{heusel2017gans} in our work.

\subsection{IP2Vec}
\label{sec:ip2vec}
\begin{figure}[h!]
	\centering
	\includegraphics[width=0.9\textwidth]{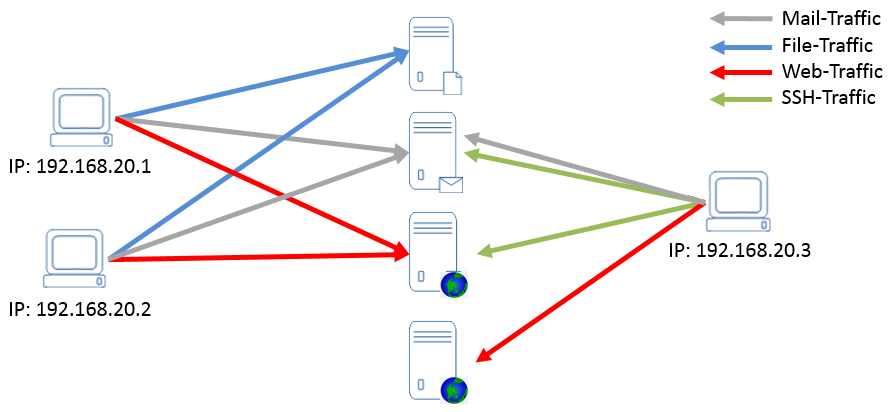}
	\caption{
		Idea of IP2Vec. 
	}
	\label{fig:generalidea}
\end{figure}

IP2Vec \cite{ring2017ip2vec} is inspired by Word2Vec \cite{mikolov2013efficient,mikolov2013distributed} and aims at transforming \textit{IP addresses} into a continuous feature space $\mathbb{R}^m$ such that standard similarity measures can be applied.
For this transformation, IP2Vec extracts available context information from flow-based network traffic.  
\textit{IP addresses} which appear frequently in similar contexts will be close to each other in the feature space $\mathbb{R}^m$.  
More precisely, similar contexts imply to IP2Vec that the devices associated to these \textit{IP addresses} establish similar network connections. 
Figure \ref{fig:generalidea} illustrates the basic idea. 

Arrows in Figure \ref{fig:generalidea} denote network connections from three \textit{IP addresses}, namely  \textit{192.168.20.1}, \textit{192.168.20.2}, and \textit{192.168.20.3}.
Colors indicate different services.
Consequently, IP2Vec leads to the following result:  
\begin{equation}
sim(192.168.20.1,192.168.20.2) > sim(192.168.20.1,192.168.20.3),
\label{frm:sim}
\end{equation}
where $sim(X,Y)$ is an arbitrary similarity function (e.g. cosine similarity) between the \textit{IP addresses} $X$ and $Y$. 
IPVec considers \textit{IP addresses} \textit{192.168.20.1} and \textit{192.168.20.2} as more similar than \textit{192.168.20.1} and \textit{192.168.20.3} because the \textit{IP addresses} \textit{192.168.20.1} and \textit{192.168.20.2} refer to the same targets and use the same services. 
In contrast to that, the \textit{IP address} \textit{192.168.20.3} targets different servers and uses different services (SSH-Connections).

\subsubsection{Model}
\begin{figure}[h!]
	\centering
	\includegraphics[width=0.9\textwidth]{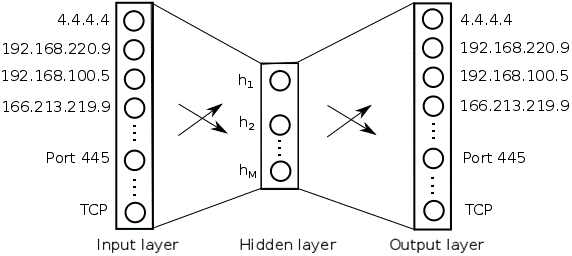}
	\caption{
		Architecture of the neural network used by IP2Vec. 
	}
	\label{fig:network}
\end{figure}

IP2Vec is based upon a fully connected neural network with a single hidden layer (see Figure \ref{fig:network}). 

The features extracted from flow-based network traffic constitute the neural network's input. 
These  features (\textit{IP addresses}, \textit{destination ports} and \textit{transport protocols}) define the input \textit{vocabulary} which contains all \textit{IP addresses}, \textit{destination ports} and \textit{transport protocols} that appear in the flow-based data set.
Since neural networks cannot be fed with categorical attributes, each value of our input \textit{vocabulary} is represented as a one-hot vector the length of which equals the size of the \textit{vocabulary}. 
Each neuron in the input and output layer is assigned a specific value of the \textit{vocabulary} (see Figure \ref{fig:network}).

Let us assume the training data set contains 100,000 different \textit{IP addresses}, 20,000 different \textit{destination ports} and 3 different \textit{transport protocols}. 
Then, the size of the one-hot vector is 120,003 and only one component is 1, while all others are 0. 
Input and output layers comprise exactly the same number of neurons which is equal to the size of the \textit{vocabulary}. 
The output layer uses a softmax classifier which indicates the probabilities for each value of the \textit{vocabulary} that it appears in the same flow (context) as the input value to the neural network. 
The softmax classifier \cite{buduma2017fundamentals} normalizes the output of all output neurons such that the sum of the outputs is 1. 
The number of neurons in the hidden layer is much smaller than the number of neurons in the input layer. 

\subsubsection{Training}
\begin{figure}[h!]
	\centering
	\includegraphics[width=0.9\textwidth]{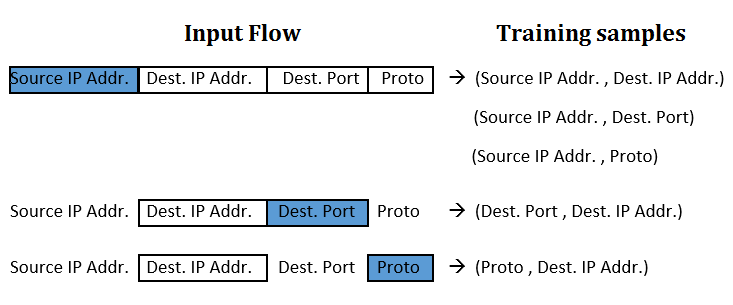}
	\caption{
		Generation of training samples in IP2Vec \cite{ring2017ip2vec}. 
		Input values are highlighted in blue color and expected output values are highlighted in black frames with white background.
	}
	\label{fig:training}
\end{figure}

The neural network is trained using captured flow-based network traffic. 
IP2Vec uses only the \textit{source IP address}, \textit{destination IP address}, \textit{destination port} and \textit{transport protocol} of flows. 
Figure \ref{fig:training} outlines the generation of training samples. 

IP2Vec generates five training samples from each flow. 
Each training sample consists of an input value and an expected output value. 
In the first step, IP2Vec selects an input value for the training sample. 
The selected input value is highlighted in blue in Figure \ref{fig:training}. 
The expected output values for the corresponding input value are highlighted through black frames with white background. 
In Figure \ref{fig:training} can be seen that IP2Vec generates three training samples where the \textit{source IP address} is the input value, one training sample where the \textit{destination port} is the input value and one training sample where the \textit{transport protocol} is the input value.

In the training process, the neural network is fed with the input value and tries to predict the probabilities of the other values from the vocabulary.  
For training samples, the probability of the concrete output value is 1 and 0 for all other values. 
In general, the output layer indicates the probabilities for each value of the input \textit{vocabulary} that it appears in the same flow as the given input value.

The network uses back-propagation for learning. 
This kind of training, however, could take a lot of time. Let us assume that the hidden layer comprises 32 neurons and the training data set encompasses one million different IP addresses and ports. 
This results in 32 million weights in each layer of the network. 
Consequently, training such a large neural network is going to be slow.
To make things worse, a huge amount of training flows is required for adjusting that many weights and for avoiding over-fitting. 
Consequently, we have to update millions of weights for millions of training samples.
Therefore, IP2Vec attempts to reduce the training time by using Negative Sampling in a similar way as Word2Vec does \cite{mikolov2013efficient}. 
In Negative Sampling, each training sample modifies only a small percentage of the weights, rather than all of them.   
More details on Negative Sampling may be found in  \cite{ring2017ip2vec} and \cite{mikolov2013distributed}.

\subsubsection{Continuous Representation of IP Addresses} 
After training the neural network, IP2Vec does not use the neural network for the task it was trained on.  
Instead, IP2Vec uses the weights of the hidden layer as $m$-dimensional vector representation of the \textit{IP addresses}.  
That means, a 32 dimensional continuous representation of each \textit{IP address}, \textit{transport protocol} and \textit{port} is obtained if the hidden layer comprises 32 neurons. 

\textit{Intuition}. 
Why does this approach work?  
If two \textit{IP addresses} refer to similar \textit{destination IP addresses}, \textit{destination ports}, and \textit{transport protocols}, then the neural network needs to output similar results for these \textit{IP addresses}. 
One way for the neural network to learn similar output values for different input values is to learn similar weights in the hidden layer of the network. 
Consequently, if two \textit{IP addresses} exhibit similar network behavior, IP2Vec attempts to learn similar weights (which are the vectors of the target feature space $\mathbb{R}^m$) in the hidden layer.

\section{Transformation Approaches}
\label{sec:approach}

This section describes three different methods to transform the heterogeneous \textit{NetFlow} data such that they may be processed by Improved Wasserstein Generative Adversarial Networks (WGAN-GP). 

\subsection{Preliminaries}
In general, we use in all three methods the same preprocessing steps for the attributes \textit{date first seen}, \textit{transport protocol}, and \textit{TCP flags} (see Table~\ref{tbl:netflow}). 

Usually, the concrete timestamp is marginal for generating realistic flow-based network data. 
Instead, many intrusion detection systems derive additional information from the timestamp like \textit{"is today a working day or weekend day"} or \textit{"does the event occur during typical working hours or at night"}. 
Therefore, we do not generate timestamps. 
Instead, we create two attributes \textit{weekday} and \textit{daytime}. 
To be precise, we extract the weekday information of flows and generate seven binary attributes \textit{isMonday}, \textit{isTuesday} and so on. 
Then, we interpret the daytime as seconds $[0,86400)$ and normalize them to the interval $[0,1]$. 
We transform the \textit{transport protocol} (see \#3 in Table \ref{tbl:netflow}) to three binary attributes, namely \textit{isTCP}, \textit{isUDP}, and \textit{isICMP}. 
The same procedure is followed for \textit{TCP flags} (see \#10 in Table \ref{tbl:netflow}) which are transformed to six binary attributes \textit{isURG}, \textit{isACK}, \textit{isPUS}, \textit{isSYN}, \textit{isRES}, and \textit{isFIN}.

\subsection{Method 1 - Numeric Transformation}
Although \textit{IP addresses} and \textit{ports} look like real numbers, they are actually  categorical. Yet, the simplest approach is to interpret them as numbers after all and treat them as continuous attributes.
We refer to this method as \textit{Numeric-based Improved Wasserstein Generative Adversarial Networks (short: N-WGAN-GP)}.  
This method transforms each octet of an \textit{IP address} to the interval [0,1], e.g. $192.168.220.14$ is transformed to four continuous attributes: 
($ip\_1$) $192/255 = 0.7529$, ($ip\_2$) $168/255 = 0.6588$, ($ip\_3$) $220/255 = 0.8627$ and ($ip\_4$) $14/255 = 0.0549$.
We do a similar procedure for \textit{ports} by dividing them through the highest port number, e.g. the \textit{source port = 80} will be transformed to one continuous attribute $80/65535=0.00122$. 

The attributes \textit{duration}, \textit{bytes} and \textit{packets} (see attributes \#2, \#8 and \#9 in Table \ref{tbl:netflow}) are normalized to the interval $[0,1]$. 
Table \ref{tbl:transform} provides examples and compares the three transformation methods.

\begin{table}[]
	\centering
	\caption{
		Preprocessing of flow-based data. 
		The first column provides the original flow attributes and examplarly values, the other columns show 
		the extracted features (column \textit{Attr.}) and the corresponding values (column \textit{Value})) for each of preprocessing method. 
	}
	\label{tbl:transform}
	
	\begin{adjustbox}{width=1.\textwidth}
		\begin{tabular}{l|ll|ll|ll}
			\hline
			& \multicolumn{2}{c}{N-WGAN-GP} & \multicolumn{2}{c}{B-WGAN-GP} & \multicolumn{2}{c}{E-WGAN-GP} \\
			Attribute / Value & Attr. & Value & Attr. & Value &  Attr. & Value   \\
			\hline
			date first seen &  isMonday & 1 & isMonday & 1 &  isMonday & 1 \\
			~~2018-05-28 11:39:23 &  isTuesday & 0 & isTuesday & 0 &  isTuesday & 0 \\
			&  isWednesday & 0 & isWednesday & 0 &  isWednesday & 0 \\
			&  isThursday & 0 & isThursday & 0 &  isThursday & 0 \\
			&  isFriday & 0 & isFriday & 0 &  isFriday & 0 \\
			&  isSaturday & 0 & isSaturday & 0 &  isSaturday & 0 \\
			&  isSunday & 0 & isSunday & 0 &  isSunday & 0 \\
			&  daytime & $\frac{41963}{86400}=0.485$ &  daytime & $\frac{41963}{86400}=0.485$ & daytime & $\frac{41963}{86400}=0.485$ \\
			\hline
			duration & norm\_dur & $\frac{1.503 - dur_{min}}{dur_{max}-dur_{min}}$  & norm\_dur & $\frac{1.503 - dur_{min}}{dur_{max}-dur_{min}}$ & dur\_1 & \multirow{3}{*}{$\left(\begin{array}{c}e_1\\...\\e_m\end{array}\right)$} \\
			~~1.503 & & & & & ... & \\
			& & & & & dur\_m & \\
			\hline
			transport protocol & isTCP & 1 & isTCP & 1 & isTCP & 1\\
			~~TCP & isUDP & 0 & isUDP & 0 & isUDP & 0 \\
			& isICMP & 0 & isICMP & 0 & isICMP & 0\\ 
			\hline
			IP address & ip\_1 & $\frac{192}{255}=0.7529$ & ip\_1 to ip\_8 & 1,1,0,0,0,0,0,0  & ip\_1 & \multirow{3}{*}{$\left(\begin{array}{c}e_1\\...\\e_m\end{array}\right)$} \\
			~~192.168.210.5 & ip\_2 & $\frac{168}{255}=0.6588$ & ip\_9 to ip\_16 & 1,0,1,0,1,0,0,0  & ... &\\
			& ip\_3 & $\frac{210}{255}=0.8627$ & ip\_17 to ip\_24 & 1,1,0,1,0,0,1,0 & ip\_m & \\
			& ip\_4 & $\frac{5}{255}=0.0196$ & ip\_25 to ip\_32 & 0,0,0,0,0,1,0,1 & \\
			\hline
			port & pt & $\frac{53872}{65535}=0.8220$ & pt\_1 to pt\_8 & 1,1,0,1,0,0,1,0 & pt\_1 & \multirow{3}{*}{$\left(\begin{array}{c}e_1\\...\\e_m\end{array}\right)$} \\
			~~53872 & & & pt\_9 to pt\_16 & 0,1,1,1,0,0,0,0 & ... & \\
			& & & & & pt\_m & \\
			\hline
			bytes & norm\_byt & $\frac{1.503 - byt_{min}}{byt_{max}-byt_{min}}$ & byt\_1 to byt\_8 & 0,0,0,0,0,0,0,0 & byt\_1 & \multirow{3}{*}{$\left(\begin{array}{c}e_1\\...\\e_m\end{array}\right)$} \\
			~~144 & & & byt\_9 to byt\_16 & 0,0,0,0,0,0,0,0  & ... & \\
			& & & byt\_17 to byt\_24 & 0,0,0,0,0,0,0,0  &byt\_m & \\
			& & & byt\_25 to byt\_32 & 1,0,0,1,0,0,0,0 & & \\
			\hline
			packets & norm\_pck & $\frac{1.503 - pck_{min}}{pck_{max}-pck_{min}}$ & pck\_1 to pck\_8 & 0,0,0,0,0,0,0,0 & pck\_1 & \multirow{3}{*}{$\left(\begin{array}{c}e_1\\...\\e_m\end{array}\right)$} \\
			~~1 & & & pck\_9 to pck\_16 & 0,0,0,0,0,0,0,0 & ... & \\
			& & & pck\_17 to pck\_24 & 0,0,0,0,0,0,0,0 & pck\_m & \\
			& & & pck\_25 to pck\_32 & 0,0,0,0,0,0,0,1 &  & \\
			\hline
			TCP flags & isURG & 0 & isURG & 0 & isURG & 0 \\ 
			~~.A..S. & isACK & 1 & isACK & 1 & isACK & 1 \\ 
			& isPSH & 0 & isPSH & 0 & isPSH & 0 \\
			& isRES & 0 & isRES & 0 & isRES & 0 \\
			& isSYN & 1  & isSYN & 1  & isSYN & 1  \\
			& isFIN & 0 & isFIN & 0 & isFIN & 0 \\
		\end{tabular}
	\end{adjustbox}
\end{table}

\subsection{Method 2 - Binary Transformation}
The second method creates several binary attributes  for \textit{IP addresses}, \textit{ports}, \textit{bytes}, and \textit{packets}.  
We refer to this method as \textit{Binary-based Improved Wasserstein Generative Adversarial Networks (short: B-WGAN-GP)}.
Each octet of an \textit{IP address} is mapped to it's 8-bit binary representation.
Consequently, \textit{IP addresses} are transformed into 32 binary attributes, e.g. 192.168.220.14 is transformed to 11000000 10101000 11011100 00001110. 
\textit{Ports} are converted to their 16-bit binary representation, e.g. the \textit{source port 80} is transformed to 00000000 01010000. 
For representing \textit{bytes} and \textit{packets}, we transform them to a binary representation as well and limit their length to 32 bit. 
The attribute \textit{duration} is normalized to the interval $[0,1]$. 
Table \ref{tbl:transform} shows an example for this transformation procedure.

\subsection{Method 3 - Embedding Transformation}
The third method transforms \textit{IP addresses}, \textit{ports}, \textit{duration}, \textit{bytes}, and \textit{packets} into so-called \textit{embeddings} in a $m$-dimensional continuous feature space $\mathbb{R}^m$ following the ideas in Section \ref{sec:ip2vec}. 
We refer to this method as \textit{Embedding-based Improved Wasserstein Generative Adversarial Networks (short: E-WGAN-GP)}. 

\begin{figure}[]
	\centering
	\includegraphics[width=1\textwidth]{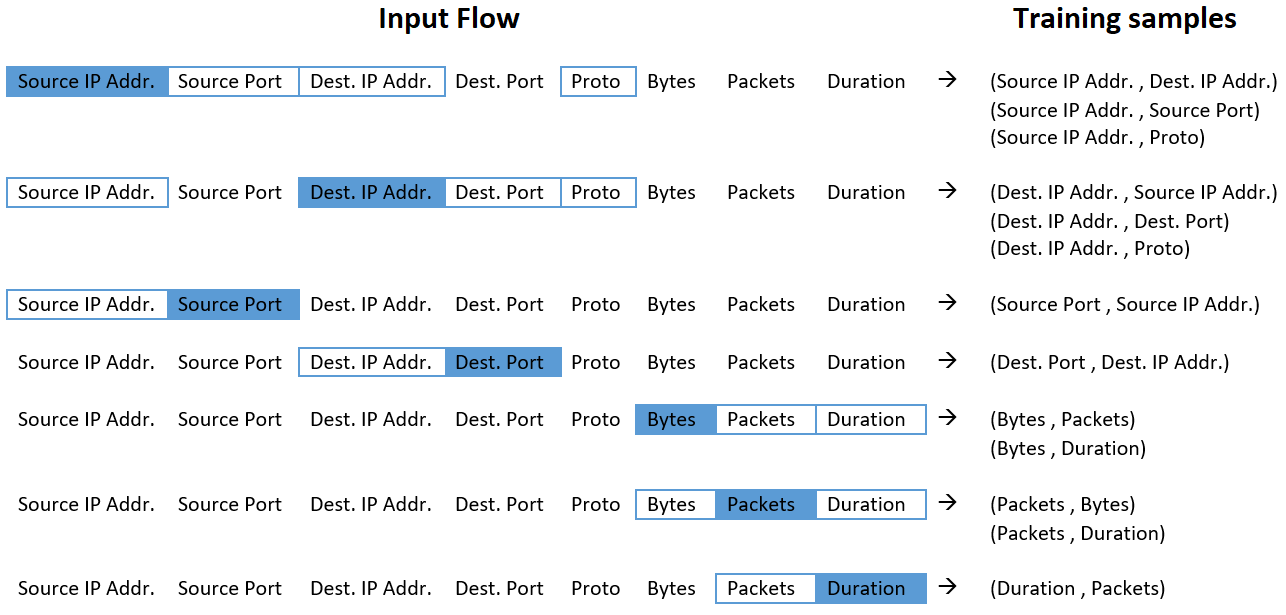}
	\caption{
		Extended generation of training samples in IP2Vec. 
		Input values are highlighted in blue color and expected output values are highlighted in black frames with white background.
	}
	\label{fig:trainingExtended}
\end{figure}

\textit{E-WGAN-GP} extends IP2Vec (see Section \ref{sec:ip2vec}) for learning embeddings not only for \textit{IP addresses}, \textit{ports}, and \textit{transport protocols}, but also for the attributes \textit{duration}, \textit{bytes}, and \textit{packets}. 
To that end, the \textit{input vocabulary} of IP2Vec is extended by the values of the latter three attributes and additional training pairs are extracted from each flow. 
Figure~\ref{fig:trainingExtended} presents the extended training sample generation. 

Each flow produces 13 training samples each of which consists of an input and an output value. 
The input values are highlighted in blue in Figure \ref{fig:trainingExtended}. 
The expected output values for the corresponding input value are highlighted through black frames on white background.
Our adapted training sample generation extracts further training samples for the attributes \textit{bytes}, \textit{packets} and \textit{duration}. 
Further, we also create training pairs with the \textit{destination IP address} as input. 
Ring et al. \cite{ring2017ip2vec} argue that it is not necessary to extract training samples with \textit{destination IP addresses} as input when working on unidirectional flows. 
Yet, in this case, IP2Vec does not learn meaningful representation for multi- and broadcast IP addresses, which only appear as \textit{destination IP addresses} in flow-based network traffic. 
Table \ref{tbl:transform} shows the result of an exemplary transformation.

\textit{E-WGAN-GP} maps flows to embeddings which need to be re-transformed  to the original space after generation.
To that end, values are replaced by the closest embeddings generated by IP2Vec. 
For instance, we calculate the cosine similarity between the generated output for the \textit{source IP address} and all existing \textit{IP address} embeddings generated by IP2Vec. 
Then, we replace the output with the \textit{IP address} with the highest similarity.

\section{Experiments}
\label{sec:experiments}
This section provides an experimental evaluation of our three approaches \textit{N-WGAN-GP}, \textit{B-WGAN-GP} and \textit{E-WGAN-GP} for synthetic flow-based network traffic generation.

\subsection{Data Set}
\label{sec:cidds}
\begin{figure}[]
	\centering
	\includegraphics[width=0.9\textwidth]{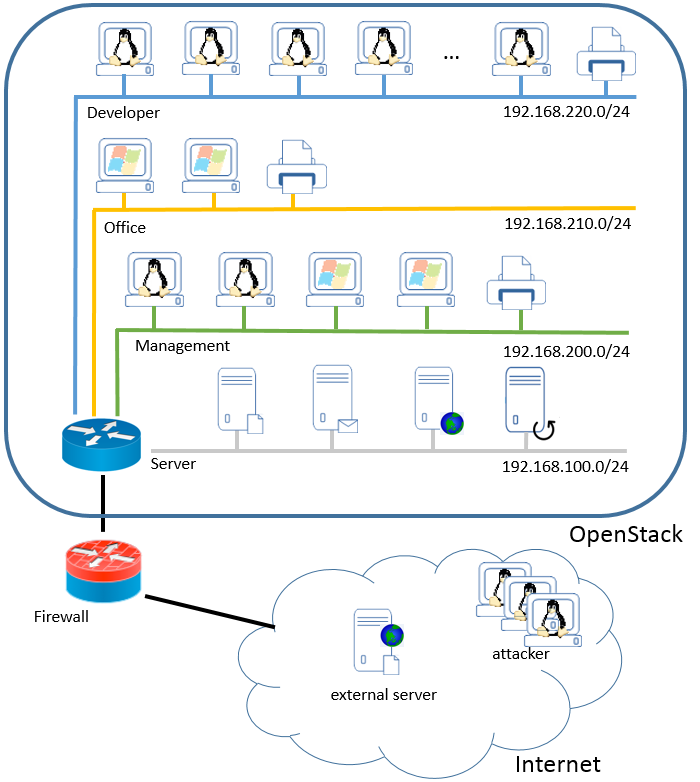}
	\caption{
		Overview of the simulated network environment from the CIDDS-001 data set~\cite{ring2017flow}.
	}
	\label{fig:cidds}
\end{figure}

We use the publicly available \textit{CIDDS-001} data set \cite{ring2017flow} which contains unidirectional flow-based network traffic as well as detailed information about the networks and \textit{IP addresses} within the data set. 
Figure \ref{fig:cidds} shows an overview of the emulated business environment of the \textit{CIDDS-001} data set. 
In essence, the \textit{CIDDS-001} data set contains four internal subnets which can be identified by their IP address ranges: a developer subnet (\textit{dev}) with exclusively Linux clients, an office subnet (\textit{off}) with exclusively Windows clients, a management subnet (\textit{mgt}) with mixed clients, and a server subnet (\textit{srv}). 
Additional knowledge facilitates the evaluation of the generated data (see Section~\ref{sec:evaluation}). 

The \textit{CIDDS-001} data set contains four weeks of network traffic. 
We consider only the network traffic which was captured at the network device within the \textit{OpenStack} environment (see Figure \ref{fig:cidds}) and divide the network traffic in two parts: \textit{week1} and \textit{week2-4}.
The first two weeks contain normal user behavior and attacks, whereas \textit{week3} and \textit{week4} contain only normal user behavior and no attacks. 
We use this kind of splitting in order to obtain a large training data set \textit{week2-4} for our generative models and simultaneously provide a reference data set \textit{week1} which contains normal and malicious network behavior. 
Overall, \textit{week2-4} contains around 22 million flows and \textit{week1} contains around 8.5 million flows. 
We consider only the TCP, UDP and ICMP flows and remove the 895 IGMP flows from the data set.

\subsection{Definition of a Baseline}
\label{sec:baseline}
As baseline for our experiments, we build a generative model which creates new flows based on the empirical probability distribution of the input data. 
The baseline estimates the probability distribution for each attribute by counting from the input data. 
New flows are generated by drawing from the empirical probability distributions. 
Each attribute is drawn independently from other attributes.

\subsection{Evaluation Methodology}
\label{sec:evaluation}

Evaluation of generative models is challenging and an open research topic:  
Borji~\cite{borji2018pros} analyzed different evaluation measures for GANs. 
Images generated with GANs are often presented to human judges and evaluated by visual comparison. 
Another well-known evaluation measure for images is the Inception Score (IS)~\cite{salimans2016improved}.  
IS classifies generated images in 1000 different classes using the Inception Net v3~\cite{szegedy2016rethinking}. 
IS, however, is not applicable in our scenario since the Inception Net v3 can only classify images, but no flow-based network traffic. 

In the IT security domain, there is neither consensus on how to evaluate network traffic generators, nor a standardized methodology~\cite{molnar2013validate}. 
Stiborek et al.~\cite{stiborek2015towards} use an anomaly score to evaluate their generated data.  
Siska et al.~\cite{siska2010} and Iannucci et al.~\cite{iannucci2017comparison} build graphs and evaluate the diversity of the generated traffic by comparing the number of nodes and edges between generated and real network traffic. 
Other flow-based network traffic generators often focus on specific aspects in their evaluation, e.g. distributions of \textit{bytes} or \textit{packets} are compared with real \textit{NetFlow} data in \cite{sommers2004self} and \cite{botta2012tool}.

Since there is no single widely accepted evaluation methodology, we use several evaluation approaches to assess the quality of the generated data from different views. 
To evaluate the diversity and distribution of the generated data, we visualize attributes (see Section \ref{sec:vis}) and compute the Euclidean distances between generated and real flow-based network data (see Section \ref{sec:euc}). 
To evaluate the quality of the content and relationships between attributes within a flow, we introduce \textit{domain knowledge checks} (see Section \ref{sec:domain}) as a new evaluation method.

\subsection{Generation of Flow-based Network Data}
Now, we evaluate the quality of the generated data by the \textit{baseline} (see Section \ref{sec:baseline}), \textit{N-WGAN-GP}, \textit{B-WGAN-GP}, and \textit{E-WGAN-GP} (see Section~\ref{sec:approach}).

\subsubsection{Parameter Configuration}
For all four approaches, we use \textit{week2-4} of the \textit{CIDDS-001} data set as training input and generate 8.5 million flows for each approach.  

We configured \textit{N-WGAN-GP}, \textit{B-WGAN-GP} and \textit{E-WGAN-GP} to use a feed-forward neural network as generator and discriminator. 
Furthermore, we used the default parameter configuration of \cite{heusel2017gans} and trained the networks for 5 epochs.
An epoch is one training iteration over the complete training data set \cite{buduma2017fundamentals}. 
Consequently, we used each flow of the training data set five times for training the neuronal networks.
We observed that a higher number of epochs neither leads to increasing quality nor reduces the loss values of the GANs.
For identifying the number of neurons in each hidden layer, we set up a small parameter study in which we varied to number of neurons from 8 to 192. 
We found that 80 neurons in each hidden layer were sufficient for \textit{B-WGAN-GP} and \textit{E-WGAN-GP}. 
For \textit{N-WGAN-GP}, we set the number of neurons in the hidden layer to 24 since the numerical representation of flows is much smaller than for \textit{B-WGAN-GP} or \textit{E-WGAN-GP}.

Additionally, we have to learn embeddings for \textit{E-WGAN-GP} in a previous step. Therefore, we configured IP2Vec to use 20 neurons in the hidden layer and trained the network like Ring et al. \cite{ring2017ip2vec} for 10 epochs. 

\subsubsection{Visualization} 
\label{sec:vis}

\begin{figure}[]
	\centering
	\includegraphics[width=0.9\textwidth]{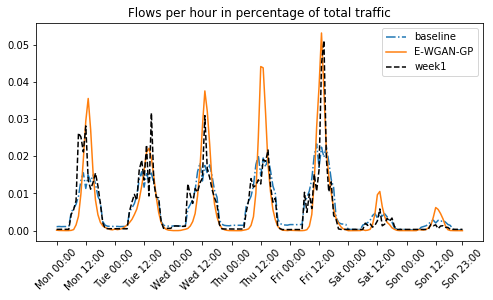}
	\caption{
		Temporal distribution of flows per hour. 
	}
	\label{fig:flowsperhour}
\end{figure}

Figure \ref{fig:flowsperhour} shows the temporal distribution of the generated flows and reference week \textit{week1}. The y-axis shows the flows per hour as a percentage of total traffic and the three lines represent the reference week (\textit{week1}), the generated data of the baseline (\textit{baseline}), and the generated data of the E-WGAN-GP approach (\textit{E-WGAN-GP}). 
Since all three transformation approaches process the attribute \textit{date first seen} in the same way, only (\textit{E-WGAN-GP}) is included for the sake of brevity. 
\textit{E-WGAN-GP} reflects the essential temporal distribution of flows. 
In the \textit{CIDDS-001} data set, the simulated users exhibit common behavior including lunch breaks and offline work which results in temporal limited network activities and a jagged curve (e.g. around 12:00 on working days).
However, the curve of \textit{E-WGAN-GP} is smoother than the curve of the original traffic \textit{week1}.

In the following, we use different visualization plots in order to get a deeper understanding of the generated data. 
\begin{figure}[h!]
	\centering
	\includegraphics[width=1.0\textwidth]{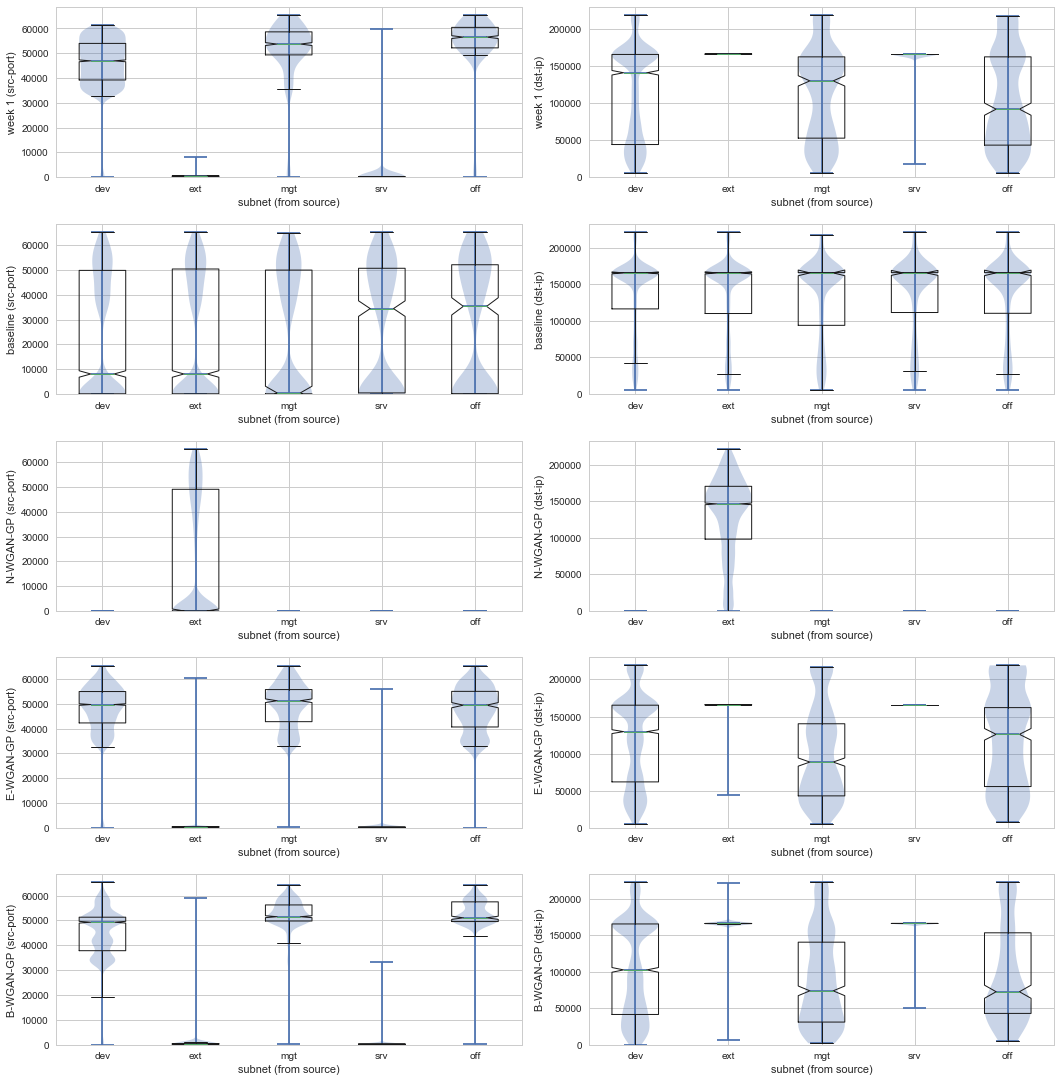}
	\caption{
		Distribution of \textit{source port} (left) and \textit{destination IP address} (right) for the subnets. 
		The rows show in order: (1) data sampled from real data (week 1) and data generated by (2) baseline, (3) N-WGAN-GP, (4) E-WGAN-GP and (5) B-WGAN-GP.
	}
	\label{fig:violins}
\end{figure}
Figure \ref{fig:violins} shows the real distributions (first row) sampled from \textit{week1} respectively generated distributions by our maximum likelihood estimator \textit{baseline} (second row) and generated distributions by our \textit{WGAN-GP} models using different data representations for each row (third to fifth row).
Each violin plot shows the data distribution of the attribute \textit{source port} (left) respectively the attribute \textit{destination IP address} (right) for the different \textit{source IP addresses} grouped by their subnet (see Section \ref{sec:cidds}).
\textit{IP addresses} from different subnets come along with different network behavior. 
For instance, \textit{IP addresses} from the \textit{mgt} subnet are typically clients which use services while \textit{IP addresses} from the \textit{srv} subnet are servers which offer services. 
This knowledge was not explicitly modeled during data generation.

We will now briefly discuss the conditional distribution of \textit{source ports} (left column). 
In the first row, we can clearly distinguish typical client-port (\textit{dev}, \textit{mgt}, \textit{off}) and server-port (\textit{ext}, \textit{srv}) distributions. 
As expected, the maximum likelihood \textit{baseline} is not able to capture the differences of the distributions depending on the subnet of the \textit{source IP address} and models a distribution which is a combination of all six subnets from the input data. 
In contrast, the \textit{B-WGAN-GP} and \textit{E-WGAN-GP} capture the conditional probability distributions for the \textit{source port} given the subnet of the \textit{source IP address} very well.

\textit{N-WGAN-GP} is incapable of representing the distributions properly: Note that only flows with external \textit{source IP addresses} are generated in the selected samples. 
In-depth analysis of the generated data suggests that numeric representations fail to match the designated subnets exactly.
As all generated data is assigned to the \textit{ext} subnet, it comes as no surprise that the distribution represents a combination of all six subnets from the input data for both \textit{source ports} (left) and \textit{destination IP addresses} (right).

For the attribute \textit{destination IP address}, the distribution is a mixture of external and internal \textit{IP addresses} for \textit{dev}, \textit{mgt} and \textit{off} subnets (see reference week \textit{week1}). 
This matches the user roles, surfing on the internet (external) as well as accessing internal services (e.g. printers). 
For external subnets, the \textit{destination IP address} has to be within the internal IP address range. 
Traffic from external sources to external targets does not run through the simulated network environment of the \textit{CIDDS-001} data set. 
Consequently, there is no flow within the \textit{CIDDS-001} data set which has a \textit{source IP address} and a \textit{destination IP address} from the \textit{ext} subnet. 
This fact can be seen for \textit{week1} in Figure \ref{fig:violins} where flows which have their origin in the \textit{ext} subnet only address a small range of \textit{destination IP addresses} which reflect the range of internal \textit{IP addresses}. 
\textit{E-WGAN-GP} and \textit{B-WGAN-GP} capture this property very well while the \textit{baseline} and \textit{N-WGAN-GP} fail to capture this property.

\subsubsection{Euclidean Distances}
\label{sec:euc}
The second evaluation compares the distribution of the generated and real flow-based network data in each attribute independently. 
Therefore, we calculate Euclidean distances between the probability distributions of the generated data and the input flow-based network data (\textit{week2-4}) in each attribute. 
We choose the Euclidean distance over the Kullback-Leibler divergence in order to avoid calculation problems where the probability of generated data is zero. 
Table \ref{tbl:resEucDis} highlights the results.
We refrain from calculating the Euclidean distance for the attribute \textit{date first seen} since exact matches of timestamps (considering seconds and milliseconds) do not make sense.  
At this point, we refer to Figure~\ref{fig:flowsperhour} which analyzes the temporal distribution of the generated timestamps. 

\begin{table}[]
	
	\caption{
		Euclidian distances between the training data (\textit{week2-4}) and the generated flow-based network traffic in each attribute. 
	}
	
	\label{tbl:resEucDis}
	\centering
	\begin{tabular}{p{10.5em}|p{3.4em}p{3.4em}p{3.4em}p{3.4em}|p{3em}}
		\noalign{\vspace{3em}}
		Attribute & \tabrotate{Baseline} & \tabrotate{N-WGAN-GP} & \tabrotate{B-WGAN-GP} & \tabrotate{E-WGAN-GP} & \tabrotate{week1} \\
		\hline
		duration 				& 0.0002 & 0.4764 & 0.4764 & 0.0525 & 0.0347\\
		transport protocol		& 0.0001 & 0.0014 & 0.0042 & 0.0015 & 0.0223 \\
		source IP address 		& 0.0003 & 0.1679 & 0.0773 & 0.0988 & 0.1409\\
		source port 			& 0.0003 & 0.5658 & 0.0453 & 0.0352 & 0.0436\\
		destination IP address 	& 0.0003 & 0.1655 & 0.0632 & 0.1272 & 0.1357 \\
		destination port 		& 0.0003 & 0.5682 & 0.0421 & 0.0327 & 0.0437\\
		bytes 					& 0.0002 & 0.5858 & 0.0391 & 0.0278 & 0.0452\\
		packets 				& 0.0004 & 1.0416 & 0.0578 & 0.0251 & 0.0437\\
		TCP flags 				& 0.0003 & 0.0217 & 0.0618 & 0.0082 & 0.0687\\
		
	\end{tabular}
	
\end{table}

Network traffic is subject to concept drift and exact reproduction of probability distributions is not desirable. 
This fact can be seen in Table \ref{tbl:resEucDis} where the Euclidean distances between the probability distributions from \textit{week1} and \textit{week2-4} of the \textit{CIDDS-001} data set are between $0.02$ and $0.14$. 
Consequently, generated network traffic should have similar Euclidean distances to the training data like the reference week \textit{week1}.
However, it should be mentioned that there is no perfect distance value $x$ which indicates the correct amount of concept drift. 
The generated data of \textit{E-WGAN-GP} tends to have similar distances to the training data (\textit{week2-4}) like the reference data set \textit{week1}. 
Table \ref{tbl:resEucDis} shows that the \textit{baseline} has the lowest distance to the training data in each attribute. 
The generated data of \textit{N-WGAN-GP} differs considerably from the training data set in some attributes. 
This is because \textit{N-WGAN-GP} often does not generate the exact values but a large number of new values. 
The binary approach \textit{B-WGAN-GP} has small distances in most attributes (except for attribute \textit{Duration}). 
This may be caused by the distribution of \textit{Duration} in the training data as most flows in the training data set have very small values in this attribute. 
Further, the normalization of the duration to interval $[0,1]$ entails that almost all flows have very low values in this attribute. \textit{N-WGAN-GP} and \textit{B-WGAN-GP} tend to generate the smallest possible duration (0.000 seconds) for all flows.

\subsubsection{Domain Knowledge Checks} 
\label{sec:domain}
We use \textit{domain knowledge checks} to evaluate the intrinsic quality of the generated data. 
To that end, we derive several properties that generated flow-based network data need to fulfill in order to be realistic.  
We use the following seven heuristics as sanity checks:  

\begin{itemize}
	\item Test 1: If the \textit{transport protocol} is UDP, then the flow must not have any TCP flags.
	\item Test 2: The \textit{CIDDS-001} data set is captured within an emulated company network. Therefore, at least one IP address (\textit{source IP address} or \textit{destination IP address}) of each flow must be internal (starting with 192.168.XXX.XXX).
	\item Test 3: If the flow describes normal user behavior and the \textit{source port} or \textit{destination port} is 80 (HTTP) or 443 (HTTPS), the \textit{transport protocol} must be \textit{TCP}.
	\item Test 4: If the flow describes normal user behavior and the \textit{source port} or \textit{destination port} is 53 (DNS), the \textit{transport protocol} must be \textit{UDP}.
	\item Test 5: If a multi- or broadcast IP address appears in the flow, it must be the \textit{destination IP address}.
	\item Test 6: If the flow represents a \textit{netbios} message (\textit{destination port} is 137 or 138), the \textit{source IP addresses} must be internal (192.168.XXX.XXX) and the \textit{destination IP address} must be an internal broadcast (192.168.XXX.255).
	\item Test 7: \textit{TCP}, \textit{UDP} and \textit{ICMP} packets have a minimum and maximum packet size. Therefore, we check the relationship between \textit{bytes} and \textit{packets} in each flow according to the following rule: 
	
	\begin{math}
	42 * packets \leq bytes \leq 65.535 * packets
	\end{math}
	
\end{itemize}
Table \ref{tbl:resDomKno} shows the results of checking the generated data against these rules.

\begin{table}[h!]
	
	\caption{
		Results of the \textit{domain knowledge checks} in percentage. 
		Higher values indicate better results.}
	
	\label{tbl:resDomKno}
	\centering
	\begin{tabular}{p{4em}|p{4em}p{4em}p{4em}p{4em}|p{4em}}
		\noalign{\vspace{3em}}
		& \tabrotate{Baseline} & \tabrotate{N-WGAN-GP} & \tabrotate{B-WGAN-GP} & \tabrotate{E-WGAN-GP} & \tabrotate{week1} \\
		\hline
		Test 1 & 14.08 & 96.46 & 97.88 & \textbf{99.77} & 100.0 \\ 
		Test 2 & 81.26 & 0.61 & 98.90 & \textbf{99.98} & 100.0 \\
		Test 3 & 86.90 & 95.45 & \textbf{99.97} & \textbf{99.97} & 100.0 \\
		Test 4 & 15.08 & 7.14 & \textbf{99.90} & 99.84 & 100.0 \\
		Test 5 & \textbf{100.0} & 25.79 & 47.13 &  99.80 & 100.0 \\ 
		Test 6 & 0.07 & 0.00 & 40.19 & \textbf{92.57} & 100.0 \\
		Test 7 & 71.26  & \textbf{100.0} & 85.32 & 99.49 & 100.0 \\
	\end{tabular}
	
\end{table}

The reference data set \textit{week1} achieves 100 percent in each test which is not surprising since the data is real flow-based network traffic which is captured in the same environment as the training data set. 
The \textit{baseline} approach does not capture dependencies between flow attributes and achieves worse results. 
This can be especially observed in Tests 1, 4, and 6. 
Since multi- and broadcast \textit{IP addresses} appear only in the attribute \textit{destination IP address}, the \textit{baseline} cannot fail Test 5 and achieves 100 percent. 

For our generative models, \textit{E-WGAN-GP} achieves the best results on average. 
The usage of embeddings leads to more meaningful similarities within categorical attributes and facilitates the learning of interrelationships. 
Embeddings, however, also reduce the possible resulting space since no new values can be generated. 
\textit{B-WGAN-GP} generates flows which achieve high accuracy in Tests 1 to 4. 
However, this approach shows weaknesses in Tests 5 and 6 where several internal relationships must be considered. 
The numerical approach \textit{N-WGAN-GP} has the lowest accuracy in the tests. 
In particular, Test 4 shows that normalization of \textit{source port} or \textit{destination port} to a single continuous attribute is inappropriate. 
Straightforward mapping of $2^{16}$ different port values to one continuous attribute leads to too many values for a good reconstruction. 
In contrast to that, the binary representation of \textit{B-WGAN-GP} leads to better results in that test.

\section{Discussion}
\label{sec:discussion}

Flow-based network traffic consists of heterogeneous data and GANs can only process continuous input values. 
To solve this problem, we analyze three methods to transform categorical to continuous attributes. The advantages and disadvantages of these approaches are discussed in the following.

\textit{N-WGAN-GP} is a straightforward numeric method but leads to unwanted similarities between categorical values which are not similar considering real data.
For instance, this transformation approach assesses the \textit{IP addresses} 192.168.220.10 and 191.168.220.10 as highly similar although the first \textit{IP address} 192.168.220.10 is private and the second \textit{IP Address} 191.168.220.10 is public. Hence, the two addresses should be ranked as fairly dissimilar.
Obviously, even small errors in the generation process can cause significant errors.
This effect can be observed in Test 2 (see Table \ref{tbl:resDomKno}) where \textit{N-WGAN-GP} has problems with the generation of private \textit{IP addresses}. Instead, this approach often generates  non-private \textit{IP addresses} such as 191.168.X.X or 192.167.X.X. 
In image generation, the original application domain of GANs, small errors do not have serious consequences.  
A brightness 191 instead of 192 in a generated pixel has nearly no effect on the image and the error is (normally) not visible for human eyes. 
Further, \textit{N-WGAN-GP} normalizes the numeric attributes \textit{bytes} and \textit{packets} to the interval $[0,1]$. 
The generated data are then de-normalized using the original training data. 
Here, we can observe that real flows often have typical byte sizes like 66 bytes which are also not exactly matched. 
This results in higher Euclidean distances in these attributes (see Table \ref{tbl:resEucDis}).  
Overall, the first method \textit{N-GAN-WP} does not seem to be suitable for generating realistic flow-based network traffic. 

\textit{B-WGAN-GP} extracts binary attributes from categorical attributes and converts numerical attributes to their binary representation. 
Using this transformation, additional structural information (e.g. subnet information) of \textit{IP addresses} can be maintained. 
Further, \textit{B-WGAN-GP} assigns larger value ranges to categorical values in the transformed space than \textit{N-WGAN-GP}. 
While \textit{N-WGAN-GP} uses a single continuous attribute to represent a \textit{source port}, \textit{B-WGAN-GP} uses 16 binary attributes for representation. 
These two aspects support \textit{B-WGAN-GP} in generating better categorical values of a flow as can be observed in the results of the \textit{domain knowledge checks} (see e.g. Test 2 and Test 4 in Table \ref{tbl:resDomKno}).
Further, Figure~\ref{fig:violins} indicates that \textit{B-WGAN-GP} captures the internal structure of the traffic very well even though it is less restricted than \textit{E-WGAN-GP} with respect to the treatment of previously unseen values. 

\textit{E-WGAN-GP} learns embeddings for \textit{IP addresses}, \textit{ports}, \textit{bytes}, \textit{packets}, and \textit{duration}. 
These embeddings are continuous vector representations and take contextual information into account.  
As a consequence, the generation of flows is less error-prone as small variations in the embedding space generally do not change the outcome in input space much.
For instance, if a GAN introduces a small error in \textit{IP address} generation, it could find the embedding of the \textit{IP address} 192.168.220.5 as nearest neighbor instead of the embedding of the expected \textit{IP address} 192.168.220.13.
Since both \textit{IP addresses} are internal clients, the error has nearly no effect. 
As a consequence, \textit{E-WGAN-GP} achieves the best results of the generative models in the evaluation. 
Yet, this approach (in contrast to \textit{N-WGAN-GP} and \textit{B-WGAN-GP}) cannot generate previously unseen values due to the embedding translation.
This is not a problem for the attributes \textit{bytes}, \textit{packets} and \textit{duration}. 
Given enough training data, embeddings for all (important) values of  \textit{bytes}, \textit{duration} and \textit{packets} are available. 
For example, consider the attribute \textit{bytes}.
We assume that the available embedding values ${b_1,b_2,b_3,...,b_{k-1},b_k}$ sufficiently cover the possible value range of the attribute \textit{bytes}.
As specific byte-values have no particular meaning, we are only interested in the magnitude of the attribute.  
Therefore, non existing values $b_x$ can be replaced with available embedding values without adversely affecting the meaning.

The situation may be different for \textit{IP addresses} and \textit{ports}. 
\textit{IP addresses} represent hosts with a distinct complex network behavior, for instance as a web server,  printer, or  Linux client. 
Generating new \textit{IP addresses} goes along with the invention of a new host with new network behavior. 
To answer the question whether the generation of new \textit{IP addresses} is necessary, the purpose needs to be considered in which the generated data shall be used later. 
If the training set comprises more than 10,000 or 100,000 different \textit{IP addresses}, there is probably no need to generate new \textit{IP addresses} for an IDS evaluation data set. 
However, this does not hold generally. 
Instead, one should ask the following two questions: (1) are there enough different \textit{IP addresses} in the training data set and (2) is there a need to generate previously unseen \textit{IP addresses}? 
If previously unseen \textit{IP addresses} are required, \textit{E-WGAN-GP} is not suitable as transformation method, otherwise \textit{E-WGAN-GP} will generate better flows than all other approaches. 

The situation for \textit{ports} is similar to \textit{IP addresses}. 
Generally, there are 65536 different \textit{ports} and most of these \textit{ports} should appear in the training data set. 
Generating new \textit{port} values is also associated with generating new behavior. 
If the training data set comprises SSH connections (Port 22) and HTTP connections (Port 80), but no FTP connections (Port 20 and 21), generators are not able to produce realistic FTP connections if they have never seen such connections.
Since the network behavior of FTP differs greatly from SSH and HTTP, it does not make much sense to generate unseen service \textit{ports}.
However, the situation is different for typical client \textit{ports}.

Generally, GANs capture the implicit conditional probability distributions very well, given that a proper data representation is chosen which is the case for \textit{E-WGAN-GP} and \textit{B-WGAN-GP} (see Figure \ref{fig:violins}).
While the visual differences between binary  and embedded data representations are subtle, the \textit{domain knowledge checks} show larger quality differences. 
Overall, this analysis suggests that \textit{E-WGAN-GP} and \textit{B-WGAN-GP} are able to generate good flow-based network traffic. 
While \textit{E-WGAN-GP} achieves better evaluation results, \textit{B-WGAN-GP} is not limited in the value range and is able to generate previously unseen values for example for \textit{IP addresses} or \textit{ports}.

\section{Related Work}
\label{sec:relatedwork}

Moln{\'a}r et al. \cite{molnar2013validate} give a comprehensive overview of network traffic generators, categorize them with respect to their purposes, and analyze the used validation measures. 
The authors conclude that there is no consensus on how to validate network traffic generators.  
Since the proposed approach aims to generate flow-based network traffic, the following overview considers primarily flow-based network traffic generators. Neither approaches which emulate computer networks and capture their network traffic like \cite{ring2017flow} or \cite{shiravi2012toward}, nor static intrusion detection data sets like DARPA 98 or KDD CUP 99 will be considered in the following. 
We categorize flow-based network traffic generators into (I) \textit{Replay Engines}, (II) \textit{Maximum Throughput Generators}, (III) \textit{Attack Generators}, and (IV) \textit{High-Level Generators}. 

\paragraph{Category (I)} 
As the name suggests, \textit{Replay Engines} use previously captured network traffic and replay the packets from it. 
Often, the aim of \textit{Replay Engines} is to consider the original inter packet time (IPT) behavior between the network packets. 
TCPReplay \cite{tcpreplay} and TCPivo \cite{feng2003tcpivo} are well-known representatives of this category. 
Since network traffic is subject to concept drift, replaying already known network traffic only makes limited sense for generating IDS evaluation data sets. 
Instead, a good network traffic generator should be able to generate new synthetic flow-based network traffic.

\paragraph{Category (II)}
\textit{Maximum Throughput Generators} usually aim to test  end-to-end network performance \cite{molnar2013validate}. 
Iperf \cite{tirumala1999iperf} is such a generator and can be used for testing bandwidth, delay jitter, and loss ratio characteristics. 
Consequently, methods from this category primarily aim at evaluating network (bandwidth) performance. 

\paragraph{Category (III)}
\textit{Attack Generators} use real network traffic as input and combine it with synthetically created attacks. 
FLAME \cite{brauckhoff2008flame} is a generator for malicious network traffic. 
The authors use rule-based approaches to inject e.g. \textit{port scan} attacks or \textit{denial of service} attacks. 
Vasilomanolakis et al. \cite{vasilomanolakis2016towards} present ID2T, a similar approach which combines real network traffic with synthetically created malicious network traffic. 
For creating malicious network traffic, the authors use rule-based scripts or manipulate parameters of the input network traffic.  
Sperotto et al. \cite{sperotto2009hidden} analyze \textit{ssh brute force} attacks on flow level and use a Hidden Markov Model to model the characteristics of them. 
However, their model generates only the number of \textit{bytes}, \textit{packets} and \textit{flows} during a typical attack scenario and does not generate complete flow-based data.

\paragraph{Category (IV)}
\textit{High-Level Generators} aim to generate new synthetic network traffic which contains realistic network connections. 
Stiborek et al. \cite{stiborek2015towards} propose three statistical methods to model host-based behavior on flow-based network level. 
The authors use real network traffic as input and extract typical inter- and intra-flow relations of host behavior.
New flow-based data is generated based on a time variant joint probability model which considers the extracted user behavior. 
Siska et al. \cite{siska2010} propose a graph-based method to generate flow-based network traffic. 
The authors use real network traffic as input and extract traffic templates. 
Traffic templates are extracted for each service port (e.g. port 80 (HTTP) or 53 (DNS)) and contain structural properties as well as the value distributions of flow attributes (e.g. log-normal distribution of transmitted \textit{bytes}). 
These traffic templates can be combined with user-defined traffic templates. 
A flow generator selects flow attributes from the traffic templates and generates new network traffic. 
Iannucci et al. \cite{iannucci2017comparison} propose PGPBA and PGSK, two synthetic flow-based network traffic generators.
Their generators are based on the graph generation algorithms \textit{Barabasi-Albert} (PGPBA) and \textit{Kronecker} (PGSK).
The authors initialize their graph-based approaches with network traffic in packet-based format. 
When generating new traffic, the authors first compute the probability of the attribute \textit{bytes}. 
All other attributes of flow-based data are calculated based on the conditional probability of the attribute \textit{bytes}. 
To evaluate the quality of their generated traffic, Iannucci et al. \cite{iannucci2017comparison} analyze the degree and pagerank distribution of their graphs to show the veracity of the generated data. 

Besides from that, GANs were recently introduced in the IT security domain. 
Yin et al. \cite{yingan} propose Bot-GAN, a framework which generates synthetic network data in order to improve botnet detection methods. 
However, their framework does not consider the generation of categorical attributes like \textit{IP addresses} and \textit{ports} which is one of the key contributions of our work. 
Hu and Tan \cite{hu2017generating} present a GAN based approach named MalGAN in order to generate synthetic malware examples which are able to bypass anomaly-based detection methods. 
Malware examples are represented as 160-dimensional binary attributes. 
Similarly, Rigaki and Garcia \cite{rigaki2016bringing} use a GAN based approach to adapt malware communication to avoid detection. 
However, they consider only three continuous attributes of the underlying network traffic in their approach.

The approach presented here does not simply replay existing network traffic like category (I). 
In fact, traffic generators from the first two categories have a different objective. 
Our approach belongs to category (IV) and generates new synthetic network traffic and is not limited to generating only malicious network traffic like category (III). 
While Siska et al. \cite{siska2010} and Iannucci et al. \cite{iannucci2017comparison} use domain knowledge to generate flows by defining conditional dependencies between flow attributes, we use GAN-based approaches which learn all dependencies between the flow attributes inherently.

\section{Summary}
\label{sec:summary}

Labeled flow-based data sets are necessary for evaluating and comparing anomaly-based intrusion detection methods. 
Evaluation data sets like DARPA~98 and KDD Cup 99 cover several attack scenarios as well as normal user behavior. 
These data sets, however, were captured at some point in time such that concept drift of network traffic causes static data sets to become obsolete sooner or later.

In this paper, we proposed three synthetic flow-based network traffic generators which are based on Improved Wasserstein GANs (WGAN-GP) \cite{gulrajani2017improved} using the two time scale update rule from \cite{heusel2017gans}. 
Our generators are initialized with real network traffic and then generate new flow-based network traffic. 
In contrast to previous high-level generators, our GAN-based approaches learn all internal dependencies between attributes inherently and no additional knowledge has to be modeled. 
Flow-based network traffic consists of heterogeneous data, but GANs can only process continuous input data. To overcome this challenge, we proposed three different methods to handle flow-based network data. 
In the first approach \textit{N-WGAN-GP}, we interpreted \textit{IP addresses} and \textit{ports} as continuous input values and normalized numeric attributes like \textit{bytes} and \textit{packets} to the interval $[0,1]$. 
In the second approach \textit{B-WGAN-GP}, we created binary attributes from categorical and numerical attributes.
For instance, we converted \textit{ports} to their 16-bit binary representation and extracted 16 binary attributes. 
\textit{B-WGAN-GP} is able to maintain more information (e.g. subnet information of \textit{IP addresses}) from the categorical input data. 
The third approach \textit{E-WGAN-GP} learns meaningful continuous representations of categorical attributes like \textit{IP addresses} using IP2Vec \cite{ring2017ip2vec}.
The preprocessing of \textit{E-WGAN-GP} is inspired from the text mining domain which also has to deal with non-continuous input values.  
Then, we generated new flow-based network traffic based on the \textit{CIDDS-001} data set~\cite{ring2017flow} in an experimental evaluation. 
Our experiments indicate that especially \textit{E-WGAN-GP} is able to generate realistic data which achieves good evaluation results. 
\textit{B-WGAN-GP} achieves similarly good results and is able to create new (unseen) values in contrast to \textit{E-WGAN-GP}. 
The quality of network data generated by \textit{N-WGAN-GP} is less convincing, which indicates that straight forward numeric transformation is not appropriate. 

Our research indicates that GANs are well suited for generating flow-based network traffic. 
We plan to extend our approach in order to generate sequences of flows instead of single flows. 
In addition, we want to work on the development of further evaluation methods.

\section*{Acknowledgments}
M.R. was supported by the BayWISS Consortium Digitization.
We gratefully acknowledge the support of NVIDIA Corporation with the donation of the Titan Xp GPU used for this research.


\section*{References}

\begin{thebibliography}{10}
	\bibitem{buczak2016survey}
	A.~L. Buczak, E.~Guven, {A Survey of Data Mining and Machine Learning Methods
		for Cyber Security Intrusion Detection}, IEEE Communications Surveys \&
	Tutorials 18~(2) (2016) 1153--1176.
	
	\bibitem{sommer2010outside}
	R.~Sommer, V.~Paxson, {Outside the Closed World: On Using Machine Learning For
		Network Intrusion Detection}, in: IEEE Symposium on Security and Privacy,
	IEEE, 2010, pp. 305--316.
	
	\bibitem{catania2012automatic}
	C.~A. Catania, C.~G. Garino, {Automatic network intrusion detection: Current
		techniques and open issues}, Computers \& Electrical Engineering 38~(5)
	(2012) 1062--1072.
	
	\bibitem{goodfellow2014generative}
	I.~Goodfellow, J.~Pouget-Abadie, M.~Mirza, B.~Xu, D.~Warde-Farley, S.~Ozair,
	A.~Courville, Y.~Bengio, {Generative Adversarial Nets}, in: Advances in
	Neural Information Processing Systems (NIPS), 2014, pp. 2672--2680.
	
	\bibitem{radford2015unsupervised}
	A.~Radford, L.~Metz, S.~Chintala, {Unsupervised Representation Learning with
		Deep Convolutional Generative Adversarial Networks}, in: International
	Conference on Learning Representations (ICLR), 2016.
	
	\bibitem{ledig2016photo}
	C.~Ledig, L.~Theis, F.~Huszár, J.~Caballero, A.~Cunningham, A.~Acosta,
	A.~Aitken, A.~Tejani, J.~Totz, Z.~Wang, W.~Shi, {Photo-Realistic Single Image
		Super-Resolution Using a Generative Adversarial Network}, in: IEEE Conference
	on Computer Vision and Pattern Recognition (CVPR), IEEE, 2017, pp. 105--114.
	
	\bibitem{isola2017image}
	P.~Isola, J.-Y. Zhu, T.~Zhou, A.~A. Efros, {Image-to-Image Translation with
		Conditional Adversarial Networks}, in: IEEE Conference on Computer Vision and
	Pattern Recognition (CVPR), IEEE, 2017, pp. 5967--5976.
	
	\bibitem{arjovsky2017wasserstein}
	M.~Arjovsky, S.~Chintala, L.~Bottou, {Wasserstein Generative Adversarial
		Networks}, in: International Conference on Machine Learning (ICML), 2017, pp.
	214--223.
	
	\bibitem{yu2017seqgan}
	L.~Yu, W.~Zhang, J.~Wang, Y.~Yu, {SeqGAN: Sequence Generative Adversarial Nets
		with Policy Gradient}, in: Conference on Artificial Intelligence (AAAI), AAAI
	Press, 2017, pp. 2852--2858.
	
	\bibitem{preuer2018fr}
	K.~Preuer, P.~Renz, T.~Unterthiner, S.~Hochreiter, G.~Klambauer,
	\href{http://arxiv.org/abs/1803.09518}{{Fr{\'e}chet ChemblNet Distance: A
			metric for generative models for molecules}}, CoRR abs/1803.09518.
	
	\bibitem{ring2017ip2vec}
	M.~Ring, D.~Landes, A.~Dallmann, A.~Hotho, {IP2Vec: Learning Similarities
		between IP Adresses}, in: Workshop on Data Mining for Cyber Security (DMCS),
	International Conference on Data Mining, IEEE, 2017, pp. 657--666.
	
	\bibitem{gulrajani2017improved}
	I.~Gulrajani, F.~Ahmed, M.~Arjovsky, V.~Dumoulin, A.~C. Courville, {Improved
		Training of Wasserstein GAN}, in: Advances in Neural Information Processing
	Systems (NIPS), 2017, pp. 5769--5779.
	
	\bibitem{heusel2017gans}
	M.~Heusel, H.~Ramsauer, T.~Unterthiner, B.~Nessler, S.~Hochreiter, {GANs
		Trained by a Two Time-Scale Update Rule Converge to a Local Nash
		Equilibrium}, in: Advances in Neural Information Processing Systems (NIPS),
	2017, pp. 6629--6640.
	
	\bibitem{ring2017flow}
	M.~Ring, S.~Wunderlich, D.~Grüdl, D.~Landes, A.~Hotho, {Flow-based benchmark
		data sets for intrusion detection}, in: European Conference on Cyber Warfare
	and Security (ECCWS), ACPI, 2017, pp. 361--369.
	
	\bibitem{claise2004cisco}
	B.~Claise, {Cisco Systems NetFlow Services Export Version 9}, RFC 3954 (2004).
	
	\bibitem{claise2008specification}
	B.~Claise, {Specification of the IP Flow Information Export ({IPFIX}) Protocol
		for the Exchange of IP Traffic Flow Information}, RFC 5101 (2008).
	
	\bibitem{han2011data}
	J.~Han, J.~Pei, M.~Kamber, {Data Mining: Concepts and Techniques}, 3rd Edition,
	Elsevier, 2011.
	
	\bibitem{beigi2014towards}
	E.~B. Beigi, H.~H. Jazi, N.~Stakhanova, A.~A. Ghorbani, {Towards Effective
		Feature Selection in Machine Learning-Based Botnet Detection Approaches}, in:
	IEEE Conference on Communications and Network Security, IEEE, 2014, pp.
	247--255.
	
	\bibitem{stevanovic2015analysis}
	M.~Stevanovic, J.~M. Pedersen, {An analysis of network traffic classification
		for botnet detection}, in: IEEE International Conference on Cyber Situational
	Awareness, Data Analytics and Assessment (CyberSA), IEEE, 2015, pp. 1--8.
	
	\bibitem{wagner2011machine}
	C.~Wagner, J.~Fran{\c{c}}ois, T.~Engel, et~al., {Machine Learning Approach for
		IP-Flow Record Anomaly Detection}, in: International Conference on Research
	in Networking, Springer, 2011, pp. 28--39.
	
	\bibitem{salakhutdinov2010efficient}
	R.~Salakhutdinov, H.~Larochelle, {Efficient learning of deep Boltzmann
		machines}, in: International Conference on Artificial Intelligence and
	Statistics, 2010, pp. 693--700.
	
	\bibitem{goodfellow2016nips}
	I.~Goodfellow, {NIPS 2016 tutorial: Generative Adversarial Networks}, arXiv
	preprint arXiv:1701.00160.
	
	\bibitem{mikolov2013efficient}
	T.~Mikolov, K.~Chen, G.~Corrado, J.~Dean, {Efficient Estimation of Word
		Representations in Vector Space}, arXiv preprint arXiv:1301.3781.
	
	\bibitem{mikolov2013distributed}
	T.~Mikolov, I.~Sutskever, K.~Chen, G.~S. Corrado, J.~Dean, {Distributed
		Representations of Words and Phrases and their Compositionality}, in:
	Advances in Neural Information Processing Systems (NIPS), 2013, pp.
	3111--3119.
	
	\bibitem{buduma2017fundamentals}
	N.~Buduma, N.~Locascio, {Fundamentals of Deep Learning: Designing
		Next-Generation Machine Intelligence Algorithms}, O'Reilly Media, 2017.
	
	\bibitem{borji2018pros}
	A.~Borji, {Pros and Cons of GAN Evaluation Measures}, arXiv preprint
	arXiv:1802.03446.
	
	\bibitem{salimans2016improved}
	T.~Salimans, I.~Goodfellow, W.~Zaremba, V.~Cheung, A.~Radford, X.~Chen,
	{Improved Techniques for Training GANs}, in: Advances in Neural Information
	Processing Systems (NIPS), 2016, pp. 2234--2242.
	
	\bibitem{szegedy2016rethinking}
	C.~Szegedy, V.~Vanhoucke, S.~Ioffe, J.~Shlens, Z.~Wojna, {Rethinking the
		Inception Architecture for Computer Vision}, in: IEEE Conference on Computer
	Vision and Pattern Recognition, 2016, pp. 2818--2826.
	
	\bibitem{molnar2013validate}
	S.~Moln{\'a}r, P.~Megyesi, G.~Szabo, {How to Validate Traffic Generators?}, in:
	IEEE International Conference on Communications Workshops (ICC), IEEE, 2013,
	pp. 1340--1344.
	
	\bibitem{stiborek2015towards}
	J.~Stiborek, M.~Reh{\'a}k, T.~Pevn{\`y}, {Towards scalable network host
		simulation}, in: International Workshop on Agents and Cybersecurity, 2015,
	pp. 27--35.
	
	\bibitem{siska2010}
	P.~Siska, M.~P. Stoecklin, A.~Kind, T.~Braun, {A Flow Trace Generator using
		Graph-based Traffic Classification Techniques}, in: International Wireless
	Communications and Mobile Computing Conference (IWCMC), ACM, 2010, pp.
	457--462.
	\newblock \href {http://dx.doi.org/10.1145/1815396.1815503}
	{\path{doi:10.1145/1815396.1815503}}.
	
	\bibitem{iannucci2017comparison}
	S.~Iannucci, H.~A. Kholidy, A.~D. Ghimire, R.~Jia, S.~Abdelwahed, I.~Banicescu,
	{A Comparison of Graph-Based Synthetic Data Generators for Benchmarking
		Next-Generation Intrusion Detection Systems}, in: IEEE International
	Conference on Cluster Computing (CLUSTER), IEEE, 2017, pp. 278--289.
	
	\bibitem{sommers2004self}
	J.~Sommers, P.~Barford, {Self-Configuring Network Traffic Generation}, in: ACM
	Internet Measurement Conference (ACM IMC), ACM, 2004, pp. 68--81.
	
	\bibitem{botta2012tool}
	A.~Botta, A.~Dainotti, A.~Pescap{\'e}, {A tool for the generation of realistic
		network workload for emerging networking scenarios}, Computer Networks
	56~(15) (2012) 3531--3547.
	
	\bibitem{shiravi2012toward}
	A.~Shiravi, H.~Shiravi, M.~Tavallaee, A.~A. Ghorbani, {Toward developing a
		systematic approach to generate benchmark datasets for intrusion detection},
	Computers \& Security 31~(3) (2012) 357--374.
	
	\bibitem{tcpreplay}
	A.~Turner, \href{http://tcpreplay.synfin.net/}{{Tcpreplay}}, (Date last
	accessed 14-June-2018).
	\url{http://tcpreplay.synfin.net/}
	
	\bibitem{feng2003tcpivo}
	W.-c. Feng, A.~Goel, A.~Bezzaz, W.-c. Feng, J.~Walpole, {TCPivo: A
		High-Performance Packet Replay Engine}, in: ACM Workshop on Models, Methods
	and Tools for Reproducible Network Research, ACM, 2003, pp. 57--64.
	
	\bibitem{tirumala1999iperf}
	D.~Jon, E.~Seth, M.~A. Bruce, P.~Jeff, P.~Kaustubh,
	\href{https://iperf.fr/}{{Iperf: The TCP/UDP bandwidth measurement tool}},
	(Date last accessed 14-June-2018).
	\url{https://iperf.fr/}
	
	\bibitem{brauckhoff2008flame}
	D.~Brauckhoff, A.~Wagner, M.~May, {FLAME: A Flow-Level Anomaly Modeling
		Engine}, in: Workshop on Cyber Security Experimentation and Test (CSET),
	USENIX Association, 2008, pp. 1:1--1:6.
	
	\bibitem{vasilomanolakis2016towards}
	E.~Vasilomanolakis, C.~G. Cordero, N.~Milanov, M.~M{\"u}hlh{\"a}user, {Towards
		the creation of synthetic, yet realistic, intrusion detection datasets}, in:
	IEEE Network Operations and Management Symposium (NOMS), IEEE, 2016, pp.
	1209--1214.
	
	\bibitem{sperotto2009hidden}
	A.~Sperotto, R.~Sadre, P.-T. de~Boer, A.~Pras, {Hidden Markov Model modeling of
		SSH brute-force attacks}, in: International Workshop on Distributed Systems:
	Operations and Management, Springer, 2009, pp. 164--176.
	
	\bibitem{yingan}
	C.~Yin, Y.~Zhu, S.~Liu, J.~Fei, H.~Zhang, {An Enhancing Framework for Botnet
		Detection Using Generative Adversarial Networks}, in: International
	Conference on Artificial Intelligence and Big Data (ICAIBD), 2018, pp.
	228--234.
	\newblock \href {http://dx.doi.org/10.1109/ICAIBD.2018.8396200}
	{\path{doi:10.1109/ICAIBD.2018.8396200}}.
	
	\bibitem{hu2017generating}
	W.~Hu, Y.~Tan, {Generating Adversarial Malware Examples for Black-Box Attacks
		Based on GAN}, arXiv preprint arXiv:1702.05983.
	
	\bibitem{rigaki2016bringing}
	M.~Rigaki, S.~Garcia, {Bringing a GAN to a Knife-fight: Adapting Malware
		Communication to Avoid Detection}, in: 1st Deep Learning and Security
	Workshop, San Francisco, USA, 2016.

\end{thebibliography}
\end{document}